\documentclass[slac_one]{revtex4}
\usepackage{graphicx}
\usepackage{fancyhdr}

\usepackage{changebar}
\usepackage{amsmath}

\begin{document}

\title{The MiniBooNE Experiment} 

%

\author{H. Ray}
\affiliation{Los Alamos National Laboratory, USA}

\begin{abstract}
Neutrino oscillations have been observed in three sectors : solar 
($\nu_e$ disappearance)~\cite{solar}, 
atmospheric ($\nu_{\mu}$ disappearance)~\cite{atm}, 
and accelerator ($\bar{\nu_{\mu}} \rightarrow \bar{\nu_e}$)~\cite{LSND}.  
The probability for two-neutrino oscillation is a function of four variables : 
two are determined by the conditions of the experiment, and two are the quantities fit for when performing an oscillation 
search ($\sin^2(2\theta)$ and $\Delta m^2$).
$\Delta \mathrm{m}^2$ is the difference in squares of the mass states of the neutrinos 
($\Delta \mathrm{m}^2_{12}$ = $\mathrm{m}^2_2$ - $\mathrm{m}^2_1$).
If the observed oscillations only occur between neutrinos in the Standard Model a summation 
law of the $\Delta \mathrm{m}^2$ is valid ($\Delta \mathrm{m}^2_{13}$ = $\Delta \mathrm{m}^2_{12}$ + 
$\Delta \mathrm{m}^2_{23}$).  The observed oscillations do not follow this 
summation law.  This implies one of the results is incorrect or there exists physics beyond the 
Standard Model.  While the solar and atmospheric results have been confirmed by several different experiments,  
the accelerator based result, from the Los Alamos LSND experiment~\cite{LSND}, has yet to be fully vetted.
The MiniBooNE experiment~\cite{MB}, 
located at Fermi National Laboratory, is designed to fully explore the LSND result.

MiniBooNE is in the final stages of performing a blind oscillation search ($\nu_{\mu} \rightarrow \nu_e$) using 
neutrino data collected through November, 2005.  A blind analysis is one in which you 
may analyze some of the information in all of the data, 
all of the information in some of the data, but not all of the information 
in all of the data.  As MiniBooNE hasn't yet opened the box, this discussion will focus 
on the different components of MiniBooNE relevant for the oscillation analysis.

\end{abstract}

\maketitle

\thispagestyle{fancy}



\section{OSCILLATION REVIEW}

The weak eigenstates 
of neutrinos are made up of a combination of mass eigenstates, analogous to mixing found in the lepton sector 

\begin{equation}
\begin{vmatrix}
\nu_e \\
\nu_{\mu}
\end{vmatrix} 
\noindent
= 
\begin{vmatrix}
 cos \theta & sin \theta  \\
-sin \theta & cos \theta
\end{vmatrix} 
\cdot
\begin{vmatrix}
\nu_1 \\
\nu_2
\end{vmatrix}
\end{equation}

For example, at the time of creation the muon neutrino is a combination of the two 
mass eigenstates : 

\begin{equation}\label{eq:numu0}
\mid\nu_{\mu}(0)> = -sin \theta \mid\nu_1> \ + \ cos \theta \mid\nu_2>
\end{equation}

To see what state the muon neutrino would be in at a later time requires the 
addition of a propagator term to each of the components :

\begin{equation}\label{eq:numut}
\mid\nu_{\mu}(t)> = -sin \theta \ e^{-iE_1 t} \ \mid\nu_1> \ + \ cos \theta \ e^{-iE_2 t} \ \mid\nu_2>
\end{equation}

The probability of oscillations between two weak eigenstates is the square of the bra of the final
weak eigenstate and the ket of the starting neutrino eigenstate :

\begin{equation}\label{eq:posc}
P_{osc} = \mid <\nu_e \mid \nu_{\mu}(t)> \mid^2
\end{equation}

The probability for oscillation may be simplified into the following equation :

\begin{equation}\label{eq:poscfinal}
P_{osc} = sin^2(2\theta) \cdot \sin^2(\frac{1.27 \cdot \Delta m^2 \cdot L}{E})
\end{equation}

An excellent discussion of the derivation of this equation may be found in Reference~\cite{osceqn}.
The probability has two terms which are constrained by the design of the experiment 
(L, the distance from the neutrino source to the detector, and E, the energy of the neutrino beam), 
and two terms which are fit for when performing a two-neutrino oscillation analysis ($\Delta \mathrm{m}^2$, 
and $\mathrm{sin^2 (2 \theta)}$, where $\theta$ is the mixing angle between the two neutrino states 
and $\Delta \mathrm{m}^2$ has been defined in the Abstract).

\begin{figure}
\centering
\includegraphics[width=0.45\textwidth]{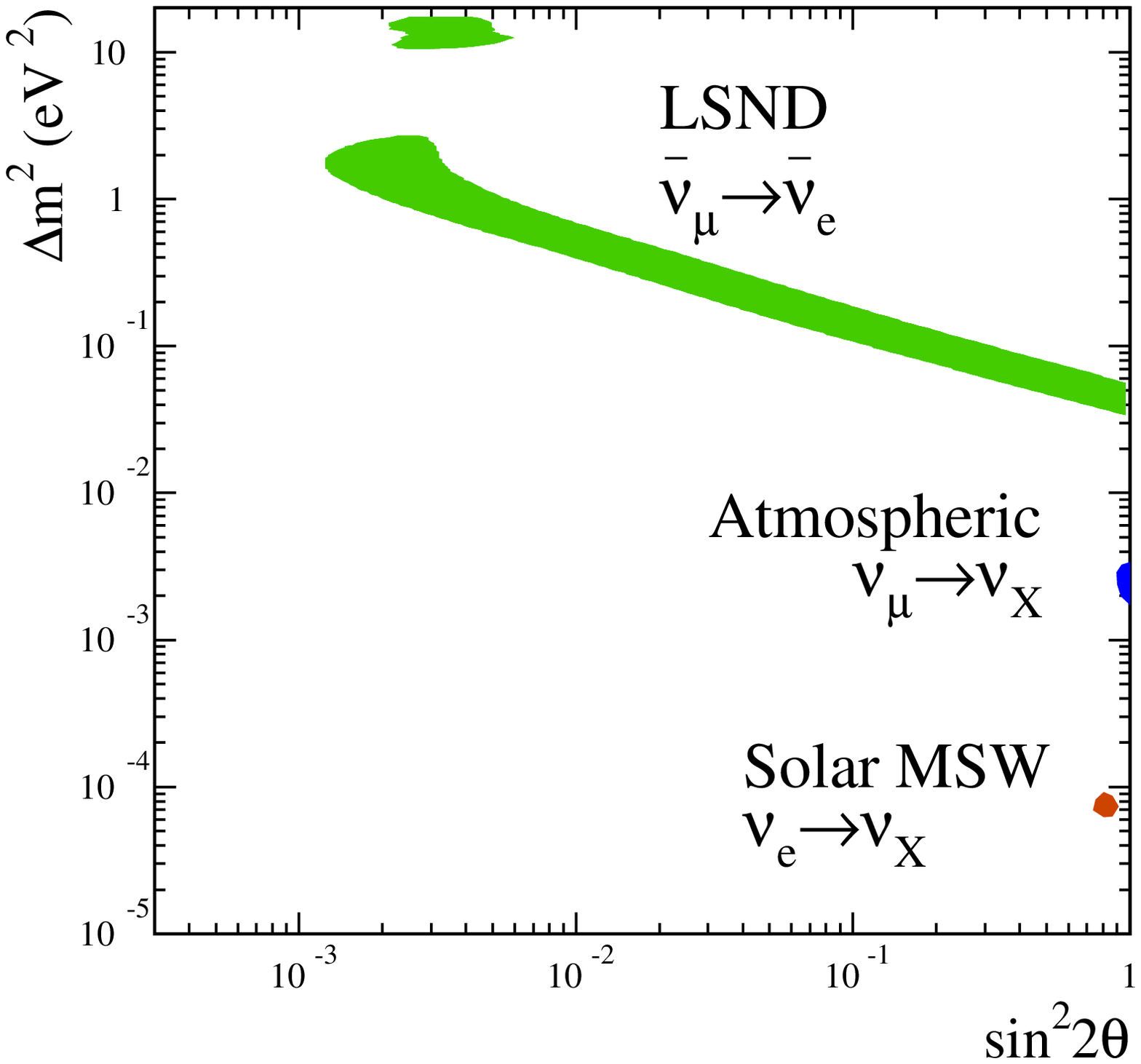}
\noindent
\includegraphics[width=0.45\textwidth]{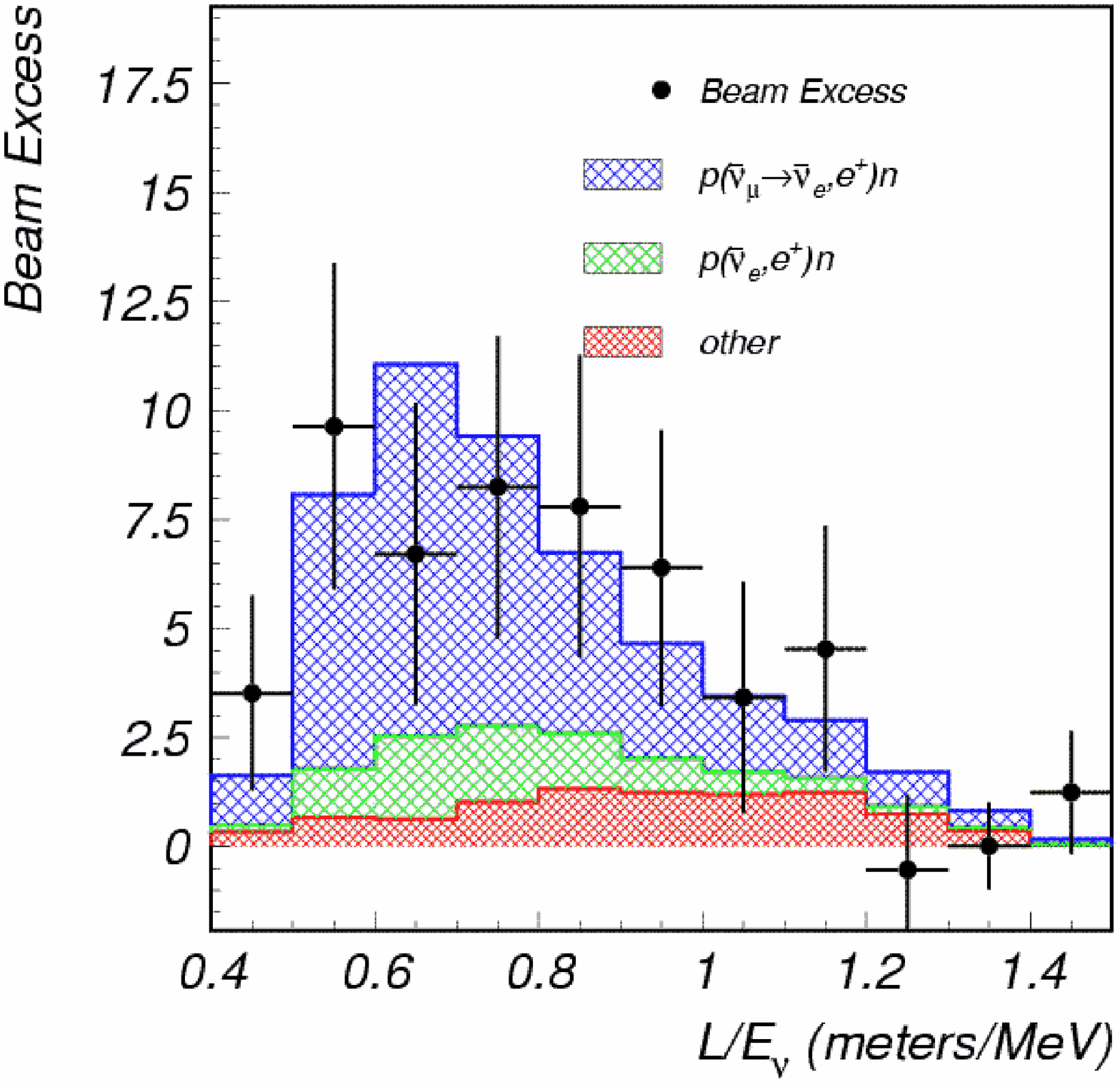}
\caption{Left : Two dimensional illustration of the current status of neutrino oscillations.  Right : Final fit result from the LSND experiment.} 
\label{fig:lsnd}
\end{figure}

Neutrino physicists illustrate the current status of neutrino oscillations using 
a two dimensional plot that is the function of the two fit parameters.  (Figure~\ref{fig:lsnd})
The oscillation results from the solar and atmospheric sectors have been observed 
and confirmed by several experiments, making the region they lie in in phase space 
very well confined.  The LSND result covers a wide swath of area.  
The $\Delta m^2$ is the mass squared difference between the two neutrino states.  
These three results represent three differences between states.  If the Standard 
Model of physics is correct and there are 3 and only 3 neutrinos, a summation 
law should exist : $\Delta \mathrm{m}^2_{13}$ = $\Delta \mathrm{m}^2_{12}$ + 
$\Delta \mathrm{m}^2_{23}$.  You can see that even at LSND's lowest allowed $\Delta m^2$ point 
the summation law does not hold.

The right portion of Figure~\ref{fig:lsnd} displays the final LSND result.  
The smaller green and red shaded areas are backgrounds; 
the blue area is the fit to the oscillation hypothesis.  To properly explore the 
LSND signal one needs an experiment that has the same experimental constraints (L/E, 
from the oscillation probability formula), higher statistics, and different 
signal signature, backgrounds, and sources of systematic errors.  This is 
MiniBooNE.

\section{THE MiniBooNE EXPERIMENT}

\subsection{How we get our neutrinos}
\begin{figure}
\centering
\includegraphics[width=0.75\textwidth]{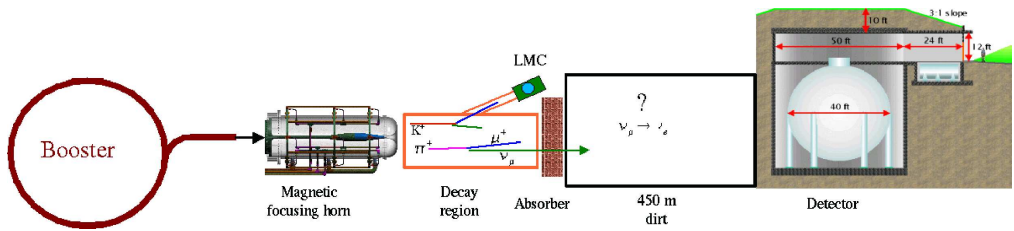}
\caption{Production and transport of the neutrino beam at MiniBooNE.} 
\label{fig:mb}
\end{figure}

MiniBooNE is located at Fermi National Laboratory, in Batavia, IL.  To produce our 
neutrino beam we start with an 8 GeV beam of protons from the Booster.  The 
proton beam enters a magnetic focusing horn where it strikes a Beryllium 
target.  (Figure~\ref{fig:mb})  We can change the polarity of the horn to direct a neutrino or 
anti-neutrino beam toward our detector.  We have collected approximately $6 \cdot 10^{20}$ protons on target 
(POT) for the neutrino oscillation result, for a total of around 600,000 neutrino events.  
In January, 2006, we switched the polarity of the horn and began running in 
anti-neutrino mode.  As of October 2006 MiniBooNE has collected $\sim 1 \cdot 10^{20}$ POT in 
anti-neutrino mode. 
 
The proton and Beryllium interactions produce a stream of charged mesons (kaons and pions).  
The mesons decay in flight into the neutrino beam seen by the detector : $K^+/\pi^+ \rightarrow \mu^+
+ \nu_{\mu}, \ \mu^+ \rightarrow e^+ + \nu_e + \bar{\nu_{\mu}}$, 
where the $\nu_{\mu}$ comprise the neutrino beam seen at MiniBooNE.  
These mesons decay in flight in our vacuum decay region.  Following the decay region is an absorber, 
put in place to stop any muons and undecayed mesons.  The neutrino beam then 
travels through approximately 450 meters of dirt before entering the MiniBooNE 
detector.

MiniBooNE is a 12.2 meter diameter sphere.  The detector is filled with pure 
mineral oil and lined with photomultiplier tubes (PMTs).  There are two regions 
of the MiniBooNE detector :  an inner light-tight region, which has a
10\% coverage in PMTs, and an optically isolated outer region, 
known as the veto region, which aids in vetoing cosmic backgrounds.

\subsection{How we detect neutrinos}

Neutrinos interact with material in the detector.  
It's the outcome of these interactions that we look for.  
Neutrinos can interact with an electron in the atomic orbit, 
the nucleus as a whole, a free proton or nucleon bound in the nucleus, or a quark.  
Starting with the lowest energy and moving to the highest energy the 
interactions neutrinos can engage in are : elastic scattering, quasi-elastic scattering, 
single pion production, and deep inelastic scattering.  

In elastic scattering the target particle is left intact and does not change its type or flavor.  
The neutrino imparts recoil energy to 
the target, which is used to observe these interactions.

In quasi-elastic scattering the neutrino interacts with the target, changes the 
target type, and emits a charged lepton.  For example consider an
electron neutrino scattering from a neutron.  Through the exchange of a W boson 
the neutrino is turned into an electron and the neutron is turned into a proton.
These are higher energy interactions; there must be enough center of mass energy 
to make the outgoing particles.

Single pion production can be broken down into resonant and coherent production.  
In resonant production the neutrino scatters from a nucleon.  A resonance of the 
nucleon is excited, and in the process of decaying back into the ground state 
the resonance emits one or more mesons.  In coherent production the neutrino 
scatters from the entire nucleus.  The nucleus does not break up, so these 
interactions require a low momentum transfer.  In coherent pion production there is no 
transfer of charge.

The highest energy neutrino interaction is deep inelastic scattering.  This 
is  scattering with very large momentum transfers, and is similar to the quasi elastic 
interactions.  Here the W boson also mediates the neutrino turning into its 
partner lepton.  However, the nucleon the neutrino scatters from is blown to bits 
due to the high momentum transfer.  The W instead interacts with the quarks in the nucleon.
The quarks shower into a variety of hadrons, dissipating the energy carried by 
the W boson.

We look for the products of these neutrino interactions in 
our detector.  The passage of charged particles through the MiniBooNE detector leaves a distinct 
mark in the form of Cerenkov light~\cite{cerenkov} and scintillation light.  Cerenkov light is 
produced when a charged particle moves through the detection medium 
with a velocity greater than the speed of light in the medium (v $>$ c/n).  
This produces an electro-magnetic shock wave, similar to a sonic boom.  
The shock wave is conical and produces a ring of light which is detected by the PMTs.
We can use Cerenkov light to measure the particle's direction and to reconstruct 
the interaction vertex.  This effect occurs immediately with the particle's 
creation and is known as a prompt light signature.

Charged particles moving through the detector also may deposit energy in the medium, 
exciting the surrounding molecules.  The de-excitation of these molecules 
produces scintillation light.  This is an isotropic, delayed light source, and 
provides no information about the track direction.  We can however use the 
PMT timing information to locate the point, or vertex, where the neutrino interaction occurred.
 
\begin{figure}
\centering
\includegraphics[width=0.50\textwidth]{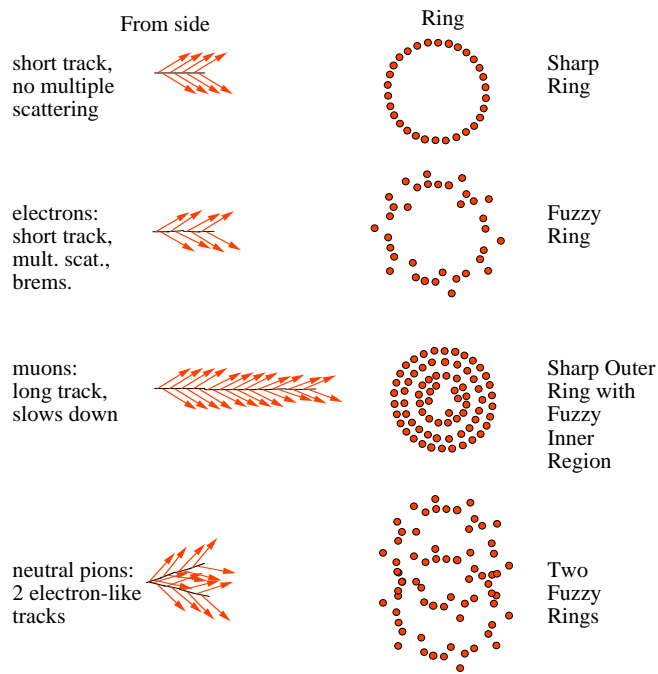}
\caption{Illustration of the Cerenkov signature seen in MiniBooNE for various particles.} 
\label{fig:sig}
\end{figure}

Figure~\ref{fig:sig} illustrates the different particle signatures we see in MiniBooNE.  
Electrons travel for only a  very short time before their velocity falls below 
the Cerenkov threshold.  They multiple scatter along the way, as well.  This 
leaves a fuzzy Cerenkov ring in the detector.  Muons tend to travel for a much longer 
distance. As they travel through the detector they lose energy, and the 
angle at which the Cerenkov light is being emitted shrinks.  The signature 
of a muon in the detector isn't one of a ring, as in the case of an electron.  
It is instead a filled in circle of light.  Neutral pions decay into two photons, which then 
pair produce.  The electrons from this pair production each create a ring in the detector.

\subsection{Components of the oscillation analysis}

MiniBooNE is performing a blind analysis.  The oscillation signal that we 
are looking for is expected to be quite small - the probability for the LSND 
oscillations was only 0.26\%!  This requires we have a very precise knowledge of 
the event rate and neutrino flux, the detector response, and the 
backgrounds to the oscillation search.  This also requires we have a well developed 
particle ID algorithm.

\subsubsection{Event Rate and Flux}
There are very few measurements of the integrated cross section for proton-Beryllium interactions, 
and none in the MiniBooNE proton energy range.  Extrapolation from the 
measurements at 6 GeV and 12 GeV to our energy range produces a flux prediction 
with very large uncertainties.  MiniBooNE is using results from two 
experiments which have just completed running to reduce the uncertainties in the 
flux prediction.  The E910 experiment~\cite{e910}, located at Brookhaven National Laboratory, 
measured pion and kaon production using a 6, 12, and 18 GeV beam of protons and 
a thin Beryllium target.  The HARP experiment at CERN~\cite{harp} took data using a 8 GeV 
proton beam and a 5, 50, and 100\% interaction length thick MiniBooNE replica 
target.  The thin target results from HARP have been completed in April 2006 and incorporated 
into the MiniBooNE Monte Carlo.  We are anxiously awaiting the thick target results.

\subsubsection{Detector Response}
A vital component of the oscillation analysis involves our ability to properly model the 
detector response.  MiniBooNE is the first neutrino experiment to use pure mineral oil as the 
detection medium.  All other neutrino experiments to date which have used 
mineral oil have doped their oil with a scintillator.  These experiments were 
looking for much lower energy neutrino interactions and they needed to boost 
the light output from the detector.  At MiniBooNE we rely on the Cerenkov light 
cone to distinguish between different particles, and thus different interactions 
and incident neutrino types.  The addition of scintillator would wash out the 
Cerenkov signature, making particle identification impossible.

As we are the first neutrino experiment to use a pure mineral oil there is no prior 
characterization of the mineral oil itself which we can use as a starting point to our simulation.  
MiniBooNE has had to perform a variety of 
tests to separate out the components of the mineral oil which create light - 
fluorescence components, scintillation light, etc.  
MiniBooNE has employed a variety of stand alone tests to characterize  the 
separate components of the mineral oil.  
While these tests have allowed us to characterize and isolate specific components 
of the mineral oil these tests provide no way to understand correlations between the components.  For that 
information we turn to our internal calibration sources.  

MiniBooNE has a muon tracker located above the tank, and several small cubes filled 
with scintillation oil located inside of the tank.  This system provides 
cosmic ray muons and their subsequent decay Michel electrons of a known 
position and direction in the tank.  These events are key to understanding 
our energy and reconstruction.  We also have 4 laser flasks which we use to measure 
tube charge and timing response, and to make sure the detector response has been 
constant as a function of time.  In addition we have an in-situ data sample, 
 the neutral current elastic sample, which provides neutrino interactions below 
the Cerenkov threshold.  These events allow us to isolate the scintillation 
components in our mineral oil and distinguish it from the fluorescence of the 
detector.

\subsubsection{Backgrounds}

Backgrounds to the oscillation analysis are broken down into two main categories : 
$\nu_{\mu}$ events which are mis-identified as $\nu_e$ events, and true $\nu_e$ 
events which are intrinsic to our beam. 

Of the $\nu_{\mu}$ events which are mis-identified as being $\nu_e$ events, $\approx$83\% are
$\pi^0$ events.  Delta decays ($\approx$7\%) and $\nu_{\mu}$ charged current quasi-elastic (CCQE) events 
($\approx$10\%) comprise the remaining $\nu_{\mu}$ backgrounds.  The $\nu_{\mu}$ CCQE 
events are well constrained using MiniBooNE data.  The $\pi^0$ background estimate is measured 
by fitting the $\pi^0$ mass peak as a function of momentum.

The intrinsic $\nu_e$ background comes from muon, kaon, and pion decays.  The muon 
and pion events are also well constrained using $\nu_{\mu}$ CCQE events.  
$\nu_e$ from kaon decays are constrained using high energy $\nu_{\mu}$ CCQE events, which are 
primarily due to kaon decays.

\subsubsection{Particle ID Algorithm}

A sensitive particle identification (PID) algorithm is required to cover the large swath of allowed 
regions in $sin^2(2\theta)$-$\Delta m^2$ phase space allowed by LSND.  We require that our PID
removes $\approx$99.9\% of the $\nu_{\mu}$ charged-current interactions, 
removes $\approx$99\% of all neutral current $\pi^0$
producing interactions, while maintaining a 30-60\% effiency for $\nu_e$ interactions.

\begin{figure}
\centering
\includegraphics[width=0.50\textwidth]{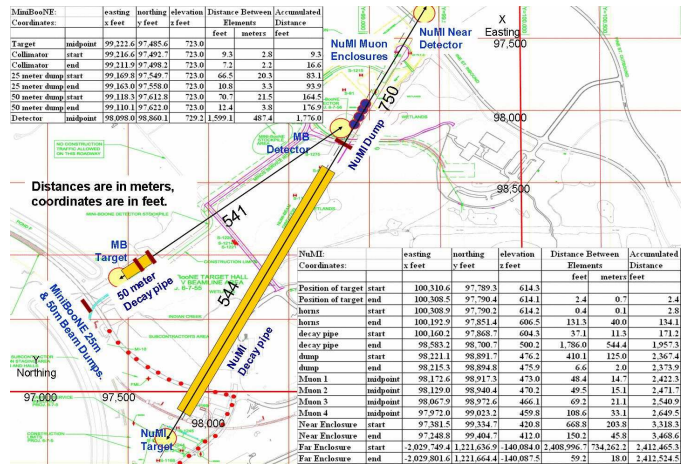}
\caption{Map of the location of the MiniBooNE detector and the NuMI target and beamline at Fermilab.} 
\label{fig:numi}
\end{figure}

To test our PID algorithm we need a sample of data which is enriched in 
$\nu_e$ events.  Our analysis is blind; we cannot use the data which contains 
$\nu_e$ interactions for calibrations as this is where the potential oscillation signal is found.  
The advent of the NuMI beamline~\cite{numi} has unintentionally allowed MiniBooNE 
to become the world's first off-axis neutrino detector.  (See Figure~\ref{fig:numi}).  
Due to this off-axis angle, the beam at MiniBooNE from NuMI is 
significantly enhanced in $\nu_e$ from $K^+$.  This provides us with a 
large sample of $\nu_e$ events to use to verify the performance of our PID algorithm.

\section{CONCLUSIONS}

All components of a blind analysis must be fully vetted and finalized before the data can be 
unblinded.  MiniBooNE is currently in the process of checking and double-checking systematic errors and 
background estimates to the oscillation analysis.  We hope to have results in the very near future.  

While the collaboration is working hard on the neutrino oscillation analysis 
our summer students have been working on analyzing the anti-neutrino data.  
We have two preliminary anti-neutrino analyses in the works : a $\bar{\nu_{\mu}}$ charged-current 
quasi-elastic free and bound proton cross section, and a neutral current $\pi^0$ analysis.

\subsection{Acknowledgments} \label{Ack}
This work was supported in part by Los Alamos National Laboratory LDRD funding.

\end{document}